\documentclass[twocolumn]{aastex63}
\usepackage{epsfig}
\usepackage{graphicx}
\usepackage{float} 
\usepackage{amsmath}
\usepackage{color}
\usepackage{amssymb}
\usepackage{amsfonts}
\usepackage{units}
\usepackage{bm}
\usepackage{longtable}
\usepackage{booktabs}
\usepackage[mathscr]{euscript}
\usepackage{soul}

\def\be{\begin{eqnarray}}
\def\ee{\end{eqnarray}}

\shorttitle{Early Optical Afterglow of GRB~240825A}
\shortauthors{Cheng et al.}  

\begin{document}

\title{Simultaneous Multiband Photometry of the Early Optical Afterglow of GRB~240825A with Mephisto}

\correspondingauthor{Yuan-Pei Yang (ypyang@ynu.edu.cn), Jinghua Zhang (zhang\_jh@ynu.edu.cn), Xiaowei Liu (x.liu@ynu.edu.cn)}

\author[0000-0001-8278-2955]{Yehao Cheng}
\affiliation{South-Western Institute for Astronomy Research, Yunnan University, Kunming, Yunnan 650504, People's Republic of China}

\author[0009-0002-7625-2653]{Yu Pan}
\affiliation{South-Western Institute for Astronomy Research, Yunnan University, Kunming, Yunnan 650504, People's Republic of China}

\author[0000-0001-6374-8313]{Yuan-Pei Yang}
\affiliation{South-Western Institute for Astronomy Research, Yunnan University, Kunming, Yunnan 650504, People's Republic of China}

\author[0000-0002-2510-6931]{Jinghua Zhang}
\affiliation{South-Western Institute for Astronomy Research, Yunnan University, Kunming, Yunnan 650504, People's Republic of China}

\author[0000-0002-8109-7152]{Guowang Du}
\affiliation{South-Western Institute for Astronomy Research, Yunnan University, Kunming, Yunnan 650504, People's Republic of China}

\author[0009-0006-1010-1325]{Yuan Fang}
\affiliation{South-Western Institute for Astronomy Research, Yunnan University, Kunming, Yunnan 650504, People's Republic of China}

\author[0000-0001-7225-2475]{Brajesh Kumar}
\affiliation{South-Western Institute for Astronomy Research, Yunnan University, Kunming, Yunnan 650504, People's Republic of China}

\author[0000-0001-5737-6445]{Helong Guo}
\affiliation{South-Western Institute for Astronomy Research, Yunnan University, Kunming, Yunnan 650504, People's Republic of China}

\author[0000-0002-8700-3671]{Xinzhong Er}
\affiliation{South-Western Institute for Astronomy Research, Yunnan University, Kunming, Yunnan 650504, People's Republic of China}

\author[0009-0000-4068-1320]{Xinlei Chen}
\affiliation{South-Western Institute for Astronomy Research, Yunnan University, Kunming, Yunnan 650504, People's Republic of China}

\author[0000-0001-5561-2010]{Chenxu Liu}
\affiliation{South-Western Institute for Astronomy Research, Yunnan University, Kunming, Yunnan 650504, People's Republic of China}

\author[0009-0005-8762-0871]{Tao Wang}
\affiliation{South-Western Institute for Astronomy Research, Yunnan University, Kunming, Yunnan 650504, People's Republic of China} 

\author[0009-0001-6598-8078]{Zhenfei Qin}
\affiliation{South-Western Institute for Astronomy Research, Yunnan University, Kunming, Yunnan 650504, People's Republic of China}

\author[0009-0003-0272-1370]{Yicheng Jin}
\affiliation{South-Western Institute for Astronomy Research, Yunnan University, Kunming, Yunnan 650504, People's Republic of China}

\author[0009-0006-5847-9271]{Xingzhu Zou}
\affiliation{South-Western Institute for Astronomy Research, Yunnan University, Kunming, Yunnan 650504, People's Republic of China}

\author[0000-0002-6107-0147]{Xuhui Han} 
\affiliation{CAS Key Laboratory of Space Astronomy and Technology, National Astronomical Observatories, Chinese Academy of Sciences, Beijing, 100101, People's Republic of China}

\author[0009-0006-1488-2587]{Pinpin Zhang} 
\affiliation{CAS Key Laboratory of Space Astronomy and Technology, National Astronomical Observatories, Chinese Academy of Sciences, Beijing, 100101, People's Republic of China}

\author[0000-0002-9422-3437]{Liping Xin}
\affiliation{CAS Key Laboratory of Space Astronomy and Technology, National Astronomical Observatories, Chinese Academy of Sciences, Beijing, 100101, People's Republic of China}

\author[0009-0001-7024-3863]{Chao Wu} 
\affiliation{CAS Key Laboratory of Space Astronomy and Technology, National Astronomical Observatories, Chinese Academy of Sciences, Beijing, 100101, People's Republic of China}

\author[0000-0001-5258-1466]{Jianhui Lian}
\affiliation{South-Western Institute for Astronomy Research, Yunnan University, Kunming, Yunnan 650504, People's Republic of China}

\author[0000-0003-0394-1298]{Xiangkun Liu}
\affiliation{South-Western Institute for Astronomy Research, Yunnan University, Kunming, Yunnan 650504, People's Republic of China}

\author[0000-0003-1295-2909]{Xiaowei Liu}
\affiliation{South-Western Institute for Astronomy Research, Yunnan University, Kunming, Yunnan 650504, People's Republic of China}

\begin{abstract}
Gamma-ray bursts (GRBs) are the most luminous transients in the universe. The interaction of the relativistic jet with the circumburst medium produces an afterglow and generates multiwavelength emission. In this work, we present simultaneous multiband photometry of GRB~240825A with the Multi-channel Photometric Survey Telescope (Mephisto) and analyze its temporal and spectral properties. The measurement began 128 seconds after the GRB trigger and continued until the fourth day when the afterglow essentially diminished and the measured brightness was close to that of the host galaxy. Based on the multiband light curves in the $uvgriz$ bands, we find that the optical flux density satisfies $F_{\nu,{\rm obs}}\propto t^{-1.34}\nu^{-2.48}$ with a spectral index of $2.48$ much larger than those of most other GRBs. To reconcile the measured much softer spectral energy distribution (SED) with that predicted by the standard afterglow model, an extra host-galaxy extinction of $E_{B-V}\sim(0.37-0.57)$ mag is required. We interpreted this excess as arising from a dense circumburst medium. We further find that the SED of the optical afterglow hardened as the afterglow decayed and the color excess $E_{B-V}$ decreased $\sim0.26$ mag from 100 seconds to 3000 seconds after the GRB trigger. Finally, we analyze the properties of the host galaxy of GRB~240825A based on data from the SDSS, PanSTARRS and HSC-SSP surveys. For a host redshift of $z=0.659$, the stellar mass and star formation rate of the host galaxy are estimated to be $\log(M_*/M_\odot)=10.0^{+0.3}_{-0.3}$ and $\log({\rm SFR}/M_{\odot}{\rm yr}^{-1})= 0.6^{+0.8}_{-3.3}$, respectively, pointing to a gas-rich, star-forming, medium-size galaxy.
\end{abstract} 


\section{Introduction}\label{intro}

Gamma-ray bursts (GRBs) are the most luminous catastrophic events in the universe. They are believed to originate from a relativistic jet powered by a newborn black hole or magnetar. Based on the duration, GRBs are generally classified as long and short GRBs \citep{Kouveliotou93}. The long GRBs have been identified as originating from core collapses of massive stars with fast rotation based on their association with broad-line Type Ic supernovae \citep[e.g.,][]{Galama98,Woosley06}, and the short GRBs are believed to originate from the merger of two compact objects including at least one neutron star (i.e., a binary neutron star or a binary consisting of a neutron star and a black hole) \citep[e.g.,][]{Eichler89,Paczynski91}.
The afterglow of a GRB usually refers to the temporal phase of emission after the prompt gamma-ray emission has ended. 
As a consequence of the interaction of relativistic jets with the circumburst medium (e.g., the interstellar medium (ISM) or the stellar wind of the GRB progenitor), the relativistic ejecta from the central engine decelerates and generates a pair of relativistic shocks: a forward shock propagating into the circumburst medium and a reverse shock penetrating the GRB ejecta itself. 
The forward and reverse shocks accelerate particles in the circumburst medium and the ejecta, respectively, and produce multiwavelength nonthermal emission of the afterglow via the synchrotron radiation \citep{Rees92,Meszaros93,Paczynski93,Katz94,Meszaros97,Sari98}.
Observations show that some afterglows of GRBs exhibit complex behaviors \citep[e.g.,][]{Akerlof99,Harrison99,Berger03} that demand sophisticated physical conditions of emission, such a medium of non-uniform density \citep{Dai98,Chevalier00}, continuous energy injection into the blastwave \citep{Dai98b,Zhang01}, joint forward- and reverse-shock emission \citep{Meszaros97,Kobayashi03,Zhang03}, etc.

The duration of the multiwavelength emission (ranging from radio to gamma-ray bands) of GRB afterglows ranges from seconds to days. Information on the emission in both temporal and spectral domains is required to study the properties of the GRB central engine and of the circumburst medium.
Very recently, GRB~240825A was triggered by the Swift Burst Alert Telescope (BAT) at 15:52:59 UT \citep{2024GCN.37274....1G} and the Fermi Gamma-ray Burst Monitor (GBM) at 15:53:00 UT \citep{2024GCN.37273....1F} on Aug. 25, 2024. 
The \texttt{Swift}--BAT light curve showed a bright peak of a complex structure of a duration of $\gtrsim$10 s, suggesting a long GRB candidate. The \texttt{Swift}--BAT count rate peaked at $\sim$1 sec after the trigger and reached $\sim1.6\times10^5$ counts/s in the band of 15-350 keV.
The UVOT on board of the \texttt{Swift} took an exposure of 150 s duration with the white filter 92 seconds after the BAT trigger, and measured the optical afterglow at a position of ${\rm RA}=344.57192$ deg and ${\rm Dec}=1.02686$ deg, with a 90\%-confidence error radius of about 0.74 arcsec \citep{2024GCN.37274....1G}.
Most multiband or multimessenger facilities all over the world performed follow-up observations of GRB~240825A, including Mephisto \citep{2024GCN.37278....1Z}, Nanshan/HMT \citep{2024GCN.37275....1J}, Skynet \citep{2024GCN.37276....1D}, AKO \citep{2024GCN.37277....1O,2024GCN.37299....1O}, Global MASTER-Net \citep{2024GCN.37279....1L,2024GCN.37283....1L}, GMG \citep{2024GCN.37280....1L,2024GCN.37306....1W}, LCO \citep{2024GCN.37287....1I}, Fermi-LAT \citep{2024GCN.37288....1D}, MASTER \citep{2024GCN.37289....1L}, Montarrenti Observatory \citep{2024GCN.37291....1L}, SVOM/C-GFT \citep{2024GCN.37292....1S} (37292, 37373), VLT/X-shooter \citep{2024GCN.37293....1M}, REM \citep{2024GCN.37295....1B}, AstroSat CZTI \citep{2024GCN.37298....1J}, MISTRAL/T193 OHP \citep{2024GCN.37300....1L}, Konus-Wind \citep{2024GCN.37302....1F}, PRIME \citep{2024GCN.37303....1G}, KAIT \citep{2024GCN.37304....1Z}, NUTTelA-TAO / BSTI \citep{2024GCN.37307....1M}, TNG \citep{2024GCN.37310....1M}, SAO RAS \citep{2024GCN.37313....1M,2024GCN.37336....1M}, ALMA \citep{2024GCN.37314....1L}, GECAM \citep{2024GCN.37315....1W}, VLA \citep{2024GCN.37322....1P}, IceCube \citep{2024GCN.37326....1I}, iTelescope \citep{2024GCN.37335....1G}, SVOM/VT \citep{2024GCN.37338....1S}, etc.
The redshift of GRB~240825A was measured at $z = 0.659$ based on the detection of multiple absorption features of Fe\,{\sc ii}, Fe\,{\sc ii}$^*$, Mn\,{\sc ii}, Mg\,{\sc ii}, Mg\,{\sc i}, Ca\,{\sc ii}, and Na\,{\sc i} of the host galaxy \citep{2024GCN.37293....1M}.

In this work, we present and analyze simultaneous multiband photometry of the optical afterglow of GRB~240825A based on the observations of the Multi-channel Photometric Survey Telescope (Mephisto). We adopt the \texttt{Swift}--BAT trigger time 15:52:59 UT on 2024 August 25 as the phase zero $T_0$. All times in the results and figures are relative to $T_0$. The paper is organized as follows. We describe our photometric observations and present the measurements of the optical afterglow of GRB~240825A in Section \ref{observation}. We analyze the physical properties of both the optical afterglow of GRB~240825A and of the host galaxy in Section \ref{analy}. The temporal and spectral evolutions of the afterglow are discussed in Section \ref{oa} and the spectral properties of the host galaxy are discussed in Section \ref{hostsed}. The results are discussed and summarized in Section \ref{conclusion}. The convention $Q_x=Q/10^x$ is adopted in cgs units unless otherwise specified.

\section{Observations of Optical afterglow of GRB~240825A}\label{observation}

The photometric observations of GRB~240825A were conducted with the Mephisto telescope. 
Mephisto is developed and operated by the South-Western Institute for Astronomy Research, Yunnan University. It is located at the Lijiang Observatory (IAU code: 044) of Yunnan Astronomical Observatories, Chinese Academy of Sciences \citep{Wang-2019RAA}, and is the first wide-field multi-channel photometric telescope of its kind worldwide. Currently, Mephisto is in the commissioning phase, equipped with two Andor Technology single-chip (e2v CCD231-C6 $6144\times 6160$ sensor) CCD cameras for the blue ($uv$) and yellow ($gr$) channels, respectively. The red channel ($iz$) is equipped with a single-chip (e2v CCD 290-99 $9216\times 9232$, 10$\mu$m sensor) camera developed at the NAOC. The spatial sampling of the red channel images, $0^{\prime\prime}.286/\text{pix}$, is better than that of the yellow and blue channel images, $0^{\prime\prime}.429/\text{pix}$. The fields of view of all those channels are $\sim$43 arcmin $\times$43 arcmin. Those will be increased to 2 sq.\,deg. once the mosaic cameras currently under development are implemented, expected to be in the spring of 2025. The three CCD cameras allow simultaneous imaging in the $u$ or $v$ (blue), $g$ or $r$ (yellow), and $i$ or $z$ (red) bands for each pointing. The wavelength ranges of the $u,v,g,r,i$, and $z$ filters are 320--365, 365--405, 480--580, 580--680, 775--900, and 900--1050 nm, respectively, with central wavelengths at 345, 385, 529, 628, 835, and 944 nm, respectively \citep{Yang24, Chen2024ApJ}. A comparison of the transmission curves of the Mephisto filters and of other systems is shown in Figure 1 of \citet{Yang24}. 

The simultaneous multiband photometric observations with Mephisto began 128 seconds after the Swift Burst Alert Telescope (BAT) triggered and located GRB~240825A. For dense sampling and better signal-to-noise ratios (SNRs), the exposure times of the three channels were adjusted to compensate the different bandwidths and efficiencies of the three channels. From 15:55:07 to 16:55:05 UTC on 2024-08-25, initial exposure times were 180, 60, and 120 seconds in the blue ($uv$), yellow ($gr$), and red ($iz$) channels, respectively. During the night, 3, 10, and 5 frames were first collected in the $vrz$ bands, followed by the same number of exposures in the $ugi$ bands. The sequence was then repeated with additional 3, 10, and 5 frames in the $ugi$ bands, and finally, the same exposures in the $vrz$ bands. In total, 6, 6, 20, 20, 10, and 10 frames were obtained in the $u,v,g,r,i$, and $z$ bands, respectively, during the first night. In the following three nights, the exposure times were increased to 300 seconds per band, and 14, 9, 14, 9, 9, and 9 frames in total were collected in the $u,v,g,r,i$, and $z$ bands, respectively. 

The raw frames were preprocessed using a dedicated pipeline developed for Mephisto (Fang et al. 2024, in preparation), including bias and dark subtraction, flat-fielding, and removal of satellite trails and cosmic rays. Astrometric calibration was performed using stars from the Gaia DR3 catalog \citep{GaiaDR3} as references. Aperture photometry was then performed using \texttt{SExtractor} \citep{Bertin1996}. The photometric calibration was carried using recalibrated Gaia BP/RP (XP) spectra with the Synthetic Photometry Method \citep{Huang2024ApJS, Xiao2023ApJS}. Synthetic magnitudes in the AB system were calculated by convolving the XP spectra with the transmission curves of the Mephisto in different bands. The XP spectra cover wavelengths from 336 to 1020 nm, which do not fully encompass wavelength ranges of the Mephisto  $u$ and $z$ bands. Therefore, we applied linear extrapolation to the XP spectra to cover missing wavelengths. The typical uncertainty in synthetic $u$ and $z$ magnitudes resulting from this extrapolation was estimates to be 0.016 mag and 0.0007 mag. We then determined the difference $\Delta m$ between the instrumental and synthetic magnitudes for good-quality non-variable stars in each frame. To mitigate the gain measurement errors, we calculated the weighted mean value of $\Delta m$ in each CCD output and applied it for the photometric calibration of targets located in that output. The weights were calculated from errors incorporating considerations from both the synthetic and instrumental magnitude uncertainties. The uncertainties in the photometric calibration were estimated to be better than 0.03, 0.01, and 0.005 mag in the $u$, $v$, and  $griz$ bands, respectively. Note that the transmission efficiency curves of Mephisto in the individual bands were calculated based on the measured efficiency curves of the individual elements in the optical path from the primary mirror to the detector.

For observations obtained in the second to fourth nights, we co-added multiple exposures in each band to enhance the SNR for better target detection. The image with the smallest median FWHM was selected as the reference for flux alignment. Flux levels of the remaining images were then scaled to that of the reference image using stars with SNRs $>$20. The same procedures as described above were applied to obtain aperture photometry in the co-added images. Photometric calibration was carried out using the zero-point values estimated from the reference image. The 3-sigma limiting magnitudes were estimated from the co-added images if the source was not clearly detected. 

We retrieved images of the host galaxy of GRB~240825A from the Panoramic Survey Telescope and Rapid Response System (PanSTARRS) \citep{Kaiser2002} archive\footnote{\url{http://ps1images.stsci.edu/cgi-bin/ps1cutouts}} and conducted aperture photometry on the stacked images in the PanSTARRS-1 $rizy$ bands. Since the source was not detected in the stacked PanSTARRS-1 $g$-band image, we estimated the 5-$\sigma$ limiting magnitude for this band. Photometric calibration was performed using the zero-point values stored in the image headers. The limiting magnitude obtained for the $g$-band is 23.70 mag, while the measured magnitudes in the $rizy$ bands are 22.61$\pm$0.19, 22.09$\pm$0.13, 22.03$\pm$0.26, and 21.71$\pm$0.21 mag, respectively, before corrected for the foreground extinction of the Milky Way. 
We also retrieved the target images from the Hyper Suprime-Cam Subaru Strategic Program \citep[HSC-SSP,][]{Aihara2018PASJ}, a wide-field multi-band imaging survey conducted with the Subaru 8.2 m telescope. The survey reaches a 5-$\sigma$ depth of 26.5, 26.5, 26.2, 25.2, and 24.4 mag in the $grizy$ bands, respectively \citep{Aihara2022PASJ}. We performed aperture photometry on the images of those bands. The images were calibrated against PanSTARRS and have a constant zero-point of 27.0 mag. The measured magnitudes in the $grizy$ bands are 23.56$\pm$0.13, 22.72$\pm$0.09, 21.93$\pm$0.06, and 21.79$\pm$0.06, and 21.54$\pm$0.06 mag, respectively, before the Galactic extinction correction.
Comparing the above values with our own measurements where we positively detected the target in the $gri$ bands on the second night and in the $griz$ bands in the third night, it appears that the GRB afterglow still has some contribution to the detected signals in the second and third nights. We also queried the Sloan Digital Sky Survey (SDSS) DR17 archive\footnote{\url{https://skyserver.sdss.org/dr17}}. The query yields host galaxy magnitudes before the Galactic extinction correction of 25.42$\pm$1.15, 23.12$\pm$0.31, 22.16$\pm$0.19, 21.82$\pm$0.23, and 21.69$\pm$0.70 mag in the SDSS $ugriz$ bands. Comparison with our data further indicates that contributions from the GRB afterglow persist in our observations on the second and third nights. However, we note that the queried SDSS magnitudes have almostly reached or exceeded the detection depths limits of the SDSS photometric survey \footnote{\url{https://www.sdss4.org/dr17/imaging/other_info/\#DepthsoftheSDSSphotometricsurvey}}. Thus, caution is warranted when interpreting these values.

\begin{figure}
\centering
\includegraphics[width = 1\linewidth, trim = 0 0 0 0, clip]{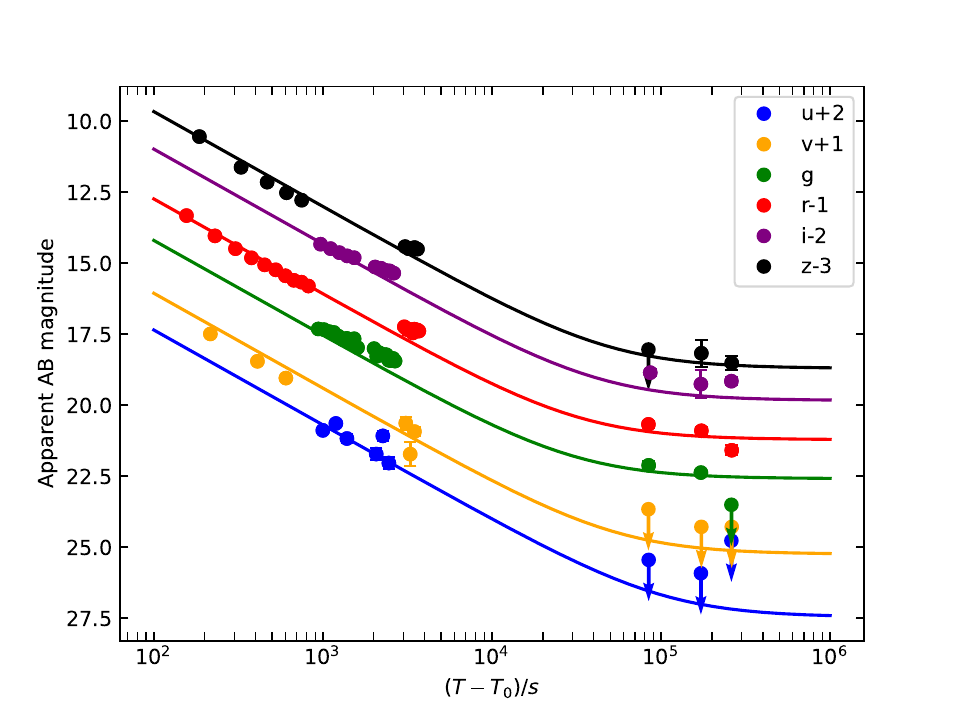}
\includegraphics[width = 1\linewidth, trim = 0 0 0 0, clip]{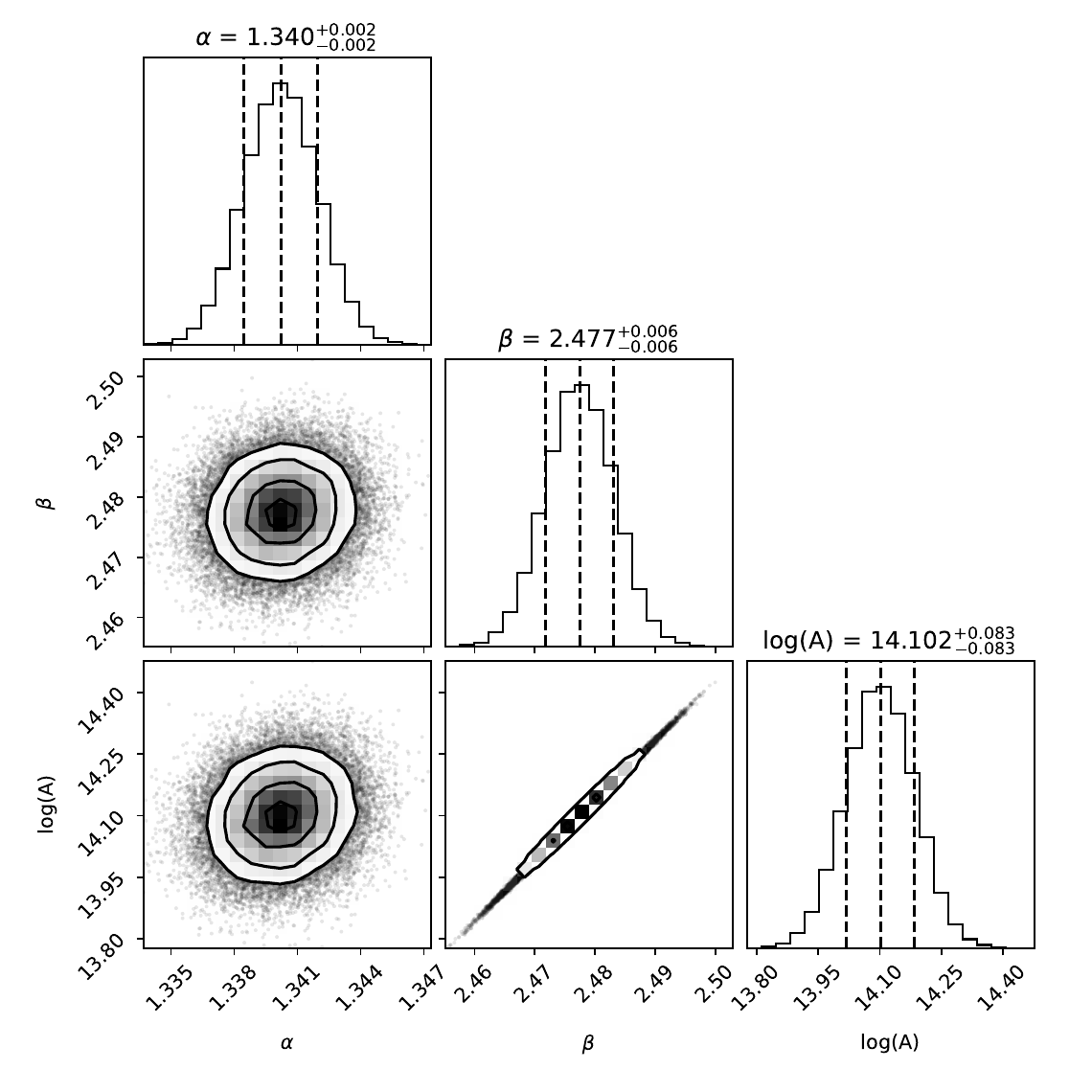}
\caption{Multiband light curves of the early optical afterglow of GRB~240825A obtained with Mephisto. Top panel: Best fit to the Mephisto multiband light curves of the optical afterglow of GRB~240825A in the $uvgriz$ bands in the first four days. The solid lines of different colors denote the best-fitting curves. The Mephisto photometric data are listed in Table \ref{tab} of the Appendix, and the times correspond to the mid times of the exposures.
Bottom panel: two-dimensional projections of the posterior probability distributions of the fitting parameters of the model given by Eq.(\ref{lc_spec}). Grey points show the data points, whereas contours show the 0.5, 1, 1.5, and 2$\sigma$ significance levels.}\label{result} 
\end{figure}

The foreground extinction from the Milky Way in the line of sight of GRB~240825A is $E_{B-V}=0.0534$ mag \citep{Schlegel98,2011schlafly}. 
Using the extinction law of \citet{Fitzpatrick1999PASP} assuming a total to selective extinction ratio of $R_V =3.1$, the extinction values in the $uvgriz$ bands are 0.264, 0.241, 0.172, 0.135, 0.086, 0.070 mag, respectively. 
In Figure \ref{result}, we present the observed light curves of the optical afterglow of GRB~240825A obtained with Mephisto in the $uvgriz$ bands, after correcting for the Galactic extinction. The data presented in Figure \ref{result} are listed in Table \ref{tab} of the Appendix along with the mid times of the measurements. Our measurements show that the early optical afterglow reaches 16.49, 14.32, 13.54 mag in the $vrz$ bands, respectively, at $\sim$130 s, and 18.89, 17.31 16.33 mag in the $ugi$ bands, respectively, at $\sim$1000 seconds after the GRB trigger.
The optical afterglow in multi-bands showed a significant power-law decay. On the fourth day after the GRB trigger, our measured magnitudes approach those of the host galaxy as given by the SDSS, PanSTARRS, and HSC-SSP observation obtained before the burst.

\section{Analysis}\label{analy}

\subsection{Properties of the Optical Afterglow of GRB~240825A}\label{oa}

A general property of the GRB afterglows is the ``multi-wavelength'' nature. As predicted by the external shock model with synchrotron or synchrotron self-Compton radiation \citep[e.g.,][]{Sari98,Sari01}, an afterglow usually covers a very wide frequency range, from the low-frequency radio to the gamma-ray range. Meanwhile, the broad-band SED of an afterglow at any instant is supposed to follow a broken power law. At a given frequency, the light curve of a GRB afterglow should also exhibit a multi-segment broken power law. 
As the blastwave is decelerated by the circumburst medium, the light curves of the afterglow at all wavelengths are expected to decay by power laws at late times (after an initial rising phase). As a result, the afterglow flux density can be characterized by,

\begin{align}
F_\nu\propto t^{-\alpha}\nu^{-\beta}\propto t^{-\alpha}\lambda^{\beta},
\end{align}
where $\nu$ is the frequency, $\lambda$ the wavelength, $\alpha$ the temporal decay index, and $\beta$ the spectral index. As the optical afterglow decays, the apparent magnitudes of the object in different bands would approach those of the host galaxy. Let the SED of the host galaxy be $F_{\nu,{\rm host}}(\nu)$, then the observed flux density can be written as,

\begin{align}
F_{\nu,{\rm obs}}(\nu, t)= A t^{-\alpha}\nu^{-\beta}+ F_{\nu,{\rm host}}(\nu),\label{lc_spec}
\end{align}
where $A$ is a constant, and $F_{\nu,{\rm host}}(\nu)$ is independent of time. The relation between the AB magnitude and the flux density is $m_\nu=-2.5\log F_{\nu,{\rm obs}}-48.6$. 

Overall, there are three general patterns of the optical afterglows of GRBs at the early time (the first few hours) as summarized by \citet{Zhang18}. 
For the first case (e.g., GRB 060418, \citet{Molinari07}), the light curve shows a smooth hump at early times and then transfers to a normal decay at late times \citep{Liang13}. This case is consistent with the external forward shock model and the hump is believed to mark the onset of the afterglow at the blastwave deceleration radius. 
The second case (e.g., GRB 990123, \citet{Akerlof99}) shows a steeper decay ($t^{-2}$ or so) early on, sometimes with a steep rising phase before the steep decay \citep{Japelj14}, and the reason is proposed to be dominated by the emission from the GRB reverse shock. 
In the third case (e.g., GRB 060729, \citet{Grupe07} and GRB 060614, \citet{Mangano07}), the light curve shows a shallow decay/plateau phase, similar to the X-ray lightcurves. Sometimes, the light curve has an early slight rise and breaks to the normal decay phase.

The optical afterglows later than a couple of hours after the GRB trigger are defined as ``late-time'' afterglows, and generally show relatively ``regular'' behaviors than those at the early time. They typically show a single power-law decay with a decay index of $\sim-1$, see the review of \citet{Zhang18}. 
If the afterglow is bright enough, a steepening break at a later time could be observable. The light curve of the late-time optical afterglow can be well fitted with a two-segment broken power law: the early part shows a normal decay of $\sim t^{-1}$, and the late part shows a steeper decay of $\sim t^{-2}$ \citep[e.g., GRB 990510 reported by][]{Harrison99}. This is consistent with the standard external forward shock model, and the temporal break is defined as the ``jet break'' \citep[e.g.,][]{Dermer00,Huang02,Zhang09}.

For the optical afterglow of GRB~240825A, based on Eq.(\ref{lc_spec}), we first apply the Bayesian inference to the measurements of GRB~240825A and derive the temporal decay index and the spectral index,

\begin{align}
\alpha_{\rm obs}=1.340\pm0.002~~~{\rm and}~~~\beta_{\rm obs}=2.477\pm0.006,
\end{align}
where $F_{\nu,{\rm host}}(\nu)$ is obtained via the SED fitting of the photometric data from the SDSS, PanSTARRS and HSC-SSP surveys, see the discussion in Section \ref{hostsed}. The data have been corrected for the extinction from the Milky Way. 
The multiband photometry of Mephisto on the first day was from 128 s to $\sim 1$ hr after the GRB trigger and corresponds to the early-time afterglow as described above. 

We define the electron spectral index as $p$ and the typical synchrotron frequencies of the standard afterglow model as $(\nu_a,\nu_m,\nu_c)$, also see the reviews of \citet{Gao13} and \citet{Zhang18}, where $\nu_a$ the synchrotron absorption frequency, $\nu_m$ the typical frequency corresponding to the minimum Lorentz factor of accelerated electrons, and $\nu_c$ the typical frequency corresponding to the cooling Lorentz factor.
Since the temporal decay index is significantly larger than one, i.e., $\alpha_{\rm obs}>1$, there are only three main possible scenarios for the standard afterglow model \citep[e.g.,][]{Gao13,Zhang18}: 

1) Case I corresponds to ISM slow cooling, $\nu_a<\min(\nu_m,\nu_c)$ with $\nu_m<\nu<\nu_c$, or $\nu_m<\nu_a<\nu_c$ with $\nu_a<\nu<\nu_c$. In this case, one has

\begin{align}
\alpha=\frac{3(p-1)}{4},~~~\beta=\frac{p-1}{2}~~~{\rm and}~~~\beta=\frac{2\alpha}{3}.\label{ab1}
\end{align}
2) Case II corresponds to wind slow cooling, $\nu_a<\min(\nu_m,\nu_c)$ with $\nu_m<\nu<\nu_c$, or $\nu_m<\nu_a<\nu_c$ with $\nu_a<\nu<\nu_c$. In this case, one has

\begin{align}
\alpha=\frac{3p-1}{4},~~~\beta=\frac{p-1}{2}~~~{\rm and}~~~\beta=\frac{2\alpha-1}{3}.\label{ab2}
\end{align}
3) Case III corresponds to ISM/wind slow/fast cooling with $\nu>\max(\nu_c,\nu_m)$. In this case, one has

\begin{align}
\alpha=\frac{3p-2}{4},~~~\beta=\frac{p}{2}~~~{\rm and}~~~\beta=\frac{2\alpha+1}{3}.\label{ab3}
\end{align}
Considering that the extinction from the circumburst medium might be significant, the measured temporal decay index $\alpha_{\rm obs}\simeq1.34$ is probably more robust than the measured spectral index $\beta_{\rm obs}\simeq2.48$. Then, for Case I, II, and III, we obtain the results of the electron spectral indexes, $p\simeq2.787\pm0.003,2.120\pm0.003,2.453\pm0.003$, respectively, and the resultant values of the spectral index (that is inferred from $\alpha_{\rm obs}$) are $\beta(\alpha_{\rm obs})\simeq0.893\pm0.001,0.560\pm0.001,1.227\pm0.001$, respectively.
We notice that our measured value of $\beta_{\rm obs}=2.477\pm0.006$ of GRB 240825A is much larger than the values of $\beta(\alpha_{\rm obs})$ that are derived and the measured values of $\beta\sim1$ from other GRBs \citep[e.g.,][]{Kann10,Kann11}.
This interesting discovery may be attributed to the simultaneous photometry of the early afterglow in wide $uvgriz$ bands of Mephisto.
The result suggests that there is a significant extinction contribution from the circumburst medium. 

\begin{figure}
\centering
\includegraphics[width = 1\linewidth, trim = 0 0 0 0, clip]{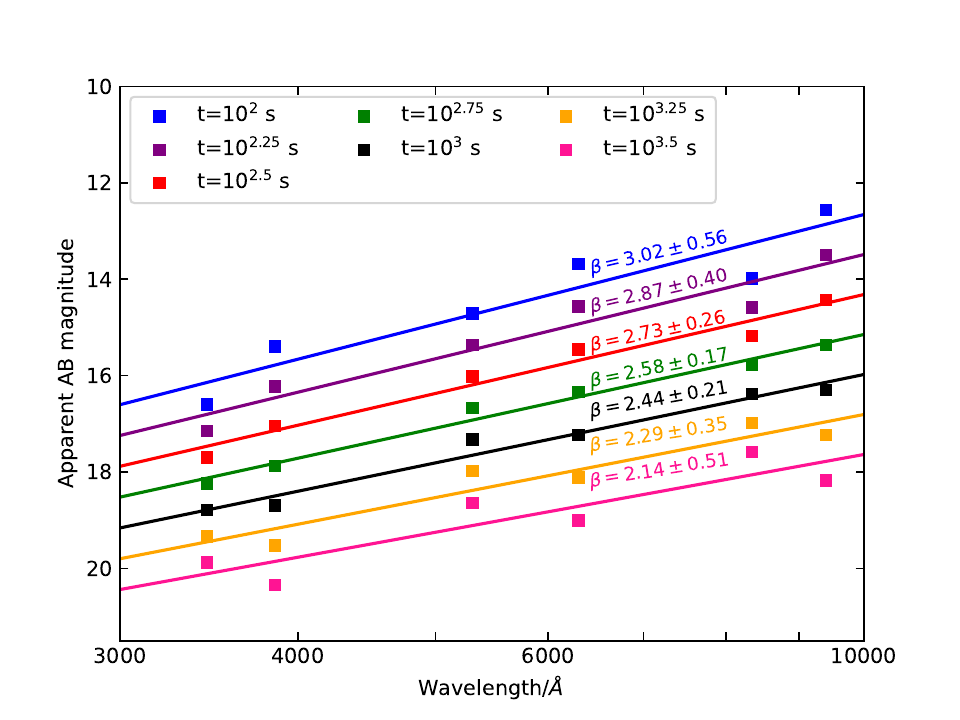}
\includegraphics[width = 1\linewidth, trim = 0 0 0 0, clip]{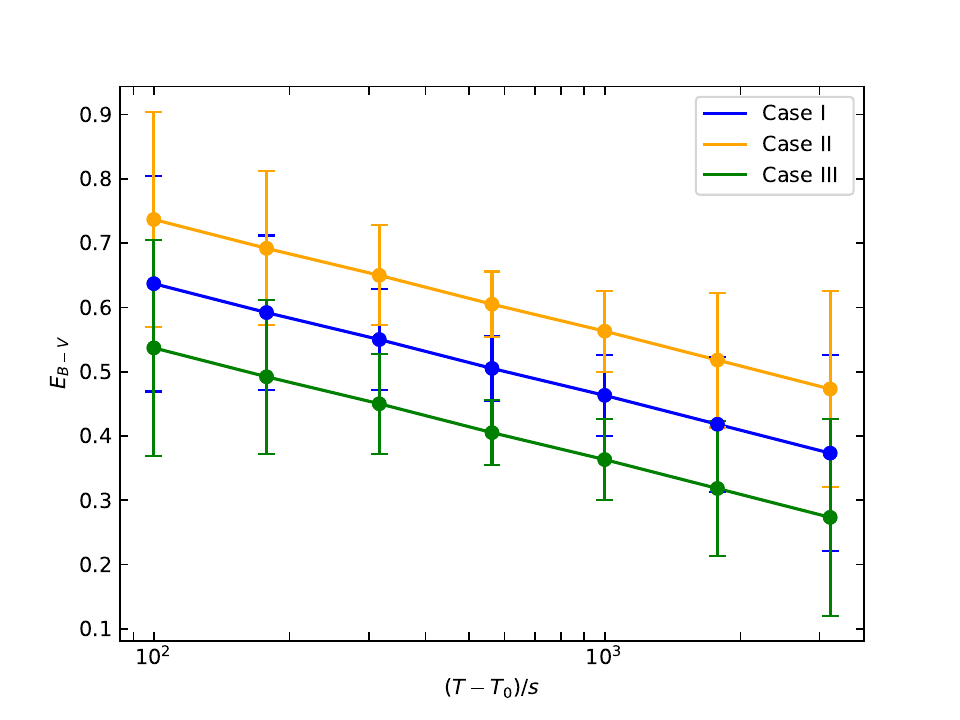}
\caption{Top panel: SEDs of the optical afterglow of GRB~240825A interpolated at different epochs (shown by different colors). One can see that the SED hardens as the afterglow decays. Bottom panel: The evolution of color excess $E_{B-V}$ that is required to reconcile the observed SED with the model prediction. Case I: ISM slow cooling, $\nu_a<\min(\nu_m,\nu_c)$ with $\nu_m<\nu<\nu_c$, or $\nu_m<\nu_a<\nu_c$ with $\nu_a<\nu<\nu_c$. Case II: wind slow cooling, $\nu_a<\min(\nu_m,\nu_c)$ with $\nu_m<\nu<\nu_c$, or $\nu_m<\nu_a<\nu_c$ with $\nu_a<\nu<\nu_c$. Case III: ISM/wind slow/fast cooling with $\nu>\max(\nu_c,\nu_m)$.
}\label{multised}
\end{figure}

Next, we discuss how much extinction is required to explain the softness of the observed SED of the optical afterglow.
Since the temporal evolution of the afterglow is independent of the extinction, the spectral index of $\beta(\alpha_{\rm obs})$ inferred from Eq.(\ref{ab1})-Eq.(\ref{ab3}) should be intrinsic, corresponding to the reference SED. On the other hand, observationally, one can measure the spectral index $\beta_{\rm obs}$ based on the observed SED. The extinction could be inferred by the difference between $\beta(\alpha_{\rm obs})$ and $\beta_{\rm obs}$ as follows.
The color between the $u$ and $z$ bands is,

\begin{align}
\Delta m_{uz}
=-2.5\log\left(\frac{F_{\nu_u}}{F_{\nu_z}}\right)=-2.5\beta\log\left(\frac{\lambda_u}{\lambda_z}\right)\simeq1.09\beta,
\end{align}
where $\lambda_u=345$ nm and $\lambda_z=944$ nm are the central wavelengths of the Mephisto $u$ and $z$ bands, respectively.
We define the observed $X$-band magnitude as $m_X$ and the intrinsic magnitude that would be observed in the absence of dust as $m_{0,X}$, then the color excess in $(u-z)$ is

\begin{align}
E_{u-z}
=\Delta m_{uz}-\Delta m_{0,uz}\simeq1.09\left[\beta_{\rm obs}-\beta(\alpha_{\rm obs})\right].
\end{align}
where $\beta(\alpha_{\rm obs})$ is given by Eq.(\ref{ab1})-Eq.(\ref{ab3}) for different cases.
Therefore, for Case I, II, and III, we have $E_{u-z}=1.73, 2.09, 1.36$ mag, respectively. 
Due to the lack of information on the circumburst medium of GRB~240825A, we assume that it has an extinction law similar to that of the Milky Way\footnote{Although the extinction profiles of GRB afterglows are usually described using that of the small Magellanic cloud (SMC) \citep[e.g.,][]{Zafar18}, in the optical band, the difference of two extinction curves of the Milky Way and of the SMC is not significant \citep[see][]{Cardelli89,Gordon03}.}. Thus, for the $u$ and $z$ bands, we have $R_u\simeq4.95$ and $R_z\simeq1.31$, according to the Galactic extinction law of \citet{Fitzpatrick1999PASP} with $R_v=3.1$. Thus, the color excess in $B-V$ could be estimated as,

\begin{align}
E_{B-V}=\frac{E_{u-z}}{R_u-R_z}.
\end{align}
In conclusion, in order to explain the observed soft SED, $E_{B-V}=0.48, 0.57, 0.37$ mag are required for Case I, II, and III, respectively. This result implies that the ISM in the host galaxy or the circumburst medium of GRB~240825A has a significant effect on the extinction to the observed afterglow. This is consistent with the general picture that long GRBs are favored to originate from star formation regions.

Next, we further check whether the level of extinction evolves during the afterglow decay. If the extinction evolves as the afterglow decays, the temporal decay index $\alpha$ would also vary in the individual bands. We use an independent temporal power-law decay to fit each light curve in the individual different bands, then measure the magnitudes at certain epochs by interpolating the fitted light curves. After deducting the contribution from the host galaxy, the SEDs of the optical afterglow at different epochs are shown in the top panel of Figure \ref{multised}. We can see that the SED hardens as the afterglow decays. 
The spectral index evolves from $\beta=3.02$ to $\beta=2.14$ from 100s to 3000s in the first day, and becomes $\beta\sim1$ in the next few days.
If the hardening of the SED is attributed to the evolution of the extinction, the color excess $E_{B-V}$ would decrease as the afterglow decay. In the bottom panel of Figure \ref{multised}, we plot the evolution of the color excess $E_{B-V}$ for Case I, II, and III, respectively. The result suggests that the color excess $E_{B-V}$ is required to decrease by $\sim0.26$ mag from 100 seconds to 3000 seconds after the GRB trigger.

Furthermore, since the Mephisto multiband light curves of GRB 248025A show a single power-law decay 128 seconds after the GRB trigger, the early observations could give a strong constraint on the onset of the afterglow at the blastwave deceleration radius. In the following discussion, we will analyze both cases of the uniform ISM and wind circumburst medium.

1) For spherically symmetric uniform medium, the mass of the swept-up material at radius $R$ is $M_{\rm sw}=4\pi m_pn_0R^3/3$, where $m_p$ is the proton mass, $n_0$ the proton density, and an ionized hydrogen gas is assumed here. The blast wave will start to undergo significant deceleration when an amount of energy comparable to the initial energy $E_0$ in the blast wave is swept up. We define the initial Lorentz factor as $\Gamma_0$. The condition $\hat\gamma\Gamma_0^2M_{\rm sw}c^2=2E_0$, where an adiabatic coefficient $\hat\gamma=4/3$ has been adopted, gives the deceleration radius \citep[e.g.,][]{Zhang18},

\begin{align}
R_{\rm dec}=\left(\frac{3E_0}{2\pi\hat\gamma\Gamma_0^2m_pc^2n_0}\right)^{1/3},
\end{align} 
and the observed deceleration time scale can be calculated through,

\begin{align}
t_{\rm dec}&\simeq0.9(1+z)\frac{R_{\rm dec}}{\Gamma_0^2c}=0.9(1+z)\left(\frac{3E_0}{2\pi\hat\gamma\Gamma_0^8m_pc^5n_0}\right)^{1/3}\nonumber\\
&\simeq186~{\rm s}~(1+z)E_{0,52}^{1/3}\Gamma_{0,2}^{-8/3}n_0^{-1/3},
\end{align}
where the coefficient of 0.9 is derived from numerical integration for a constant density medium \citep{Zhang18}. Assuming that the peak time of optical afterglow corresponds to the deceleration time (i.e., $t_{\rm dec}\lesssim128~{\rm s}$) and using the redshift of $z\simeq0.659$ \citep{2024GCN.37293....1M}, one can derive the initial Lorentz factor of the fireball for the ISM model,

\begin{align}
\Gamma_0&\simeq0.9^{3/8}\left[\frac{3E_0(1+z)^3}{2\pi\hat\gamma m_pc^5n_0t_{\rm dec}^3}\right]^{1/8}\nonumber\\
&\simeq126(1+z)^{3/8}E_{0,52}^{1/8}n_0^{-1/8}t_{\rm dec,2}^{-3/8}\gtrsim139E_{0,52}^{1/8}n_0^{-1/8}.
\end{align}

2) For the wind circumburst medium with a constant mass loss rate of $\dot M$ and a wind speed of $v_w$, the wind density is $\rho=\dot M/4\pi v_w r^2=A_wr^{-2}$, where $A_w\equiv\dot M/4\pi v_w$. For an impulsive fireball, the deceleration dynamics gives,

\begin{align}
R_{\rm dec}=\frac{E_0}{2\pi\hat\gamma\Gamma_0^2c^2A_w},
\end{align}
and the deceleration time is,

\begin{align}
t_{\rm dec}&\simeq1.3(1+z)\frac{R_{\rm dec}}{\Gamma_0^2c}=1.3(1+z)\frac{E_0}{2\pi\hat\gamma\Gamma_0^4c^3A_w}\nonumber\\
&\simeq5.7~{\rm s}(1+z)E_{0,52}A_{w,11}^{-1}\Gamma_{0,2}^{-4},
\end{align}
where the coefficient of 1.3 is derived from numerical integration for a wind density medium \citep{Zhang18}. For $t_{\rm dec}\lesssim128~{\rm s}$ and $z\simeq0.659$, one can derive the initial Lorentz factor of the fireball for the wind model,

\begin{align}
\Gamma_0&\simeq1.3^{1/4}\left[\frac{E_0(1+z)}{2\pi\hat\gamma c^3A_wt_{\rm dec}}\right]^{1/4}\nonumber\\
&\simeq49(1+z)^{1/4}E_{0,52}^{1/4}A_{w,11}^{-1/4}t_{\rm dec,2}^{-1/4}\gtrsim 52E_{0,52}^{1/4}A_{w,11}^{-1/4}.
\end{align}
In summary, for both the ISM and the wind scenarios, the lower limits of the initial Lorentz factor of the GRB ejecta are constrained to be $\Gamma_0\gtrsim139E_{0,52}^{1/8}n_0^{-1/8}$ and $\Gamma_0\gtrsim52E_{0,52}^{1/4}A_{w,11}^{-1/4}$, respectively.

\subsection{Properties of the Host Galaxy of GRB~240825A}\label{hostsed}

\begin{figure}
\centering
\includegraphics[width = 1\linewidth, trim = 0 0 0 0, clip]{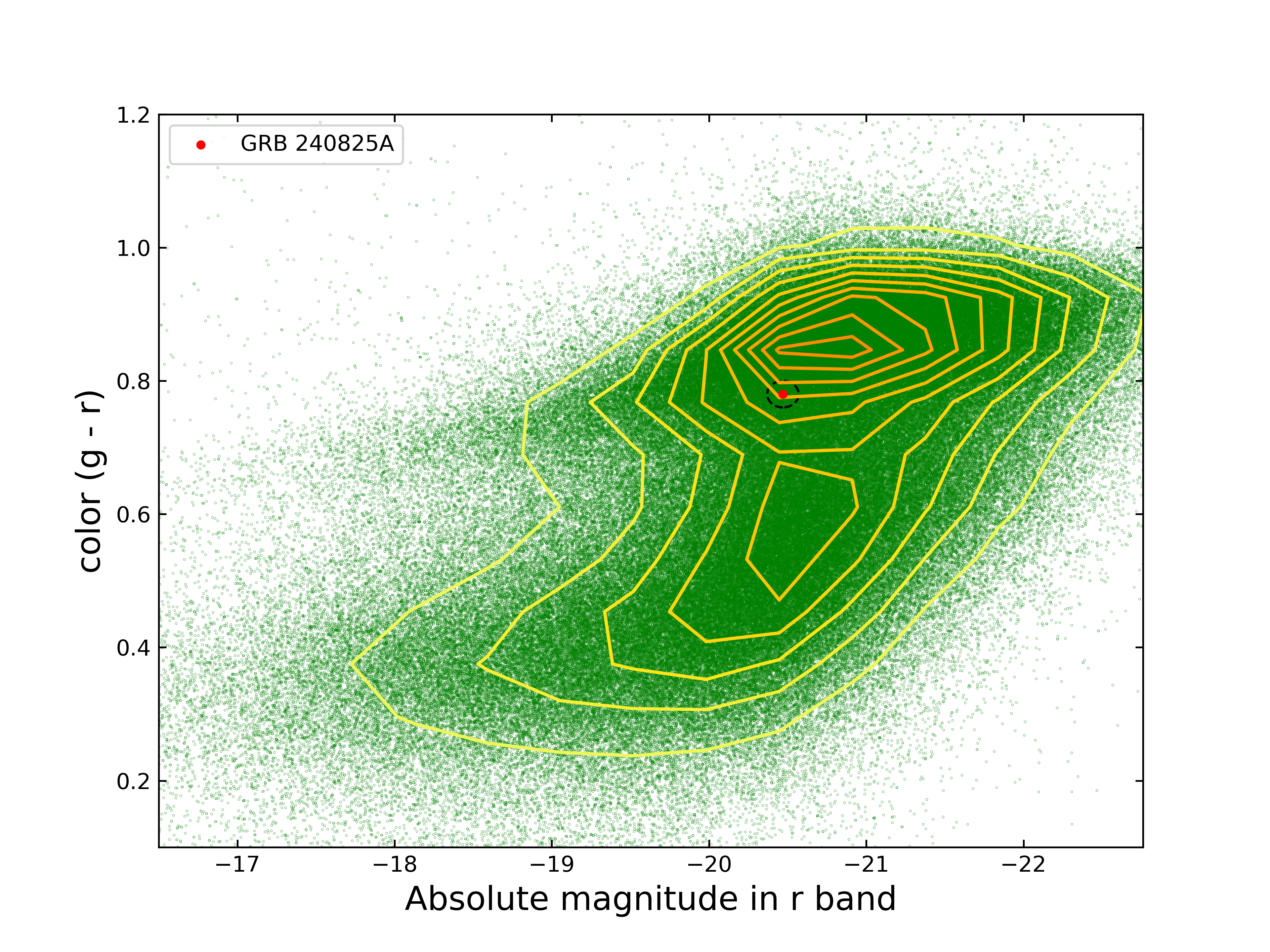}
\caption{Color $(g-r)$--magnitude (absolute AB magnitude in $r$ band) diagram of galaxies including the host galaxy of GRB~240825A. The host galaxy of GRB~240825A with $M_g=-19.69$ mag and $M_r=-20.47$ mag, as derived from the photometry of the HSC-SSP, is the red point. The sample of other galaxies is selected from SDSS DR18, shown as the green points. 
To characterize the properties of the GRB host galaxy, we select a region (the dashed circle) near the GRB host with a horizontal (magnitude) radius of 0.1 mag and a vertical (color) radius of 0.02 mag, which includes 1925 galaxies, and calculate the mean properties of the galaxies in this region. }\label{host} 
\end{figure}

In this section, we analyze the properties of the host galaxy of GRB~240825A.
We first post the HSC-SSP photometric data of the host galaxy in the color $(g-r)$ -- magnitude (the absolute AB magnitude in $r$ band) diagram of galaxies, as shown in Figure \ref{host}. 
The sample of galaxies is selected from the SDSS DR18\footnote{\href{https://skyserver.sdss.org/CasJobs/}{https://skyserver.sdss.org/CasJobs/}}. The photometric data of the host galaxy is from HSC-SSP \citep{Aihara2018PASJ}, and the absolute magnitudes in $r$ and $g$ bands are $M_r=-20.47$ mag and $M_g=-19.69$ mag after corrected for the Galactic extinction and assuming $z=0.659$ \citep{2024GCN.37293....1M}. We chose a region near the GRB host with a horizontal (magnitude) radius of 0.1 mag and a vertical (color) radius of 0.02 mag. The region includes 1925 galaxies. The average star formation rate and stellar mass of galaxies in this region are respectively of $\log({\rm SFR}/M_\odot~{\rm yr^{-1}})=-0.8\pm0.7$ and $\log(M_*/M_\odot)=10.34\pm0.08$. The result suggests that the host galaxy of GRB 240825A ranks moderately amongst galaxies sampled by the SDSS survey. 

We next discuss whether the extra extinction of $E_{B-V}\sim(0.37-0.57)$ mag derived in Section \ref{oa} originates from the ISM of the host galaxy. Assuming that the host galaxy has an extinction property similar to that of the Milky Way, then the measured color of $(g-r)=0.78$ mag of the host galaxy after corrected for a value of $-(R_g-R_r)E_{B-V}$, where $R_g=3.7$ and $R_r=2.6$ for the HSC-SSP filters, leads to the intrinsic color of the host galaxy of $(g-r)_0\simeq(0.15-0.37)$ mag. 
The magnitude of the host galaxy in the $r$ band needs to be corrected for a value of $-R_rE_{B-V}$, yielding an intrinsic absolute magnitude in $r$ band of $M_{0,r}\simeq-(21.4-21.9)$ mag. 
In conclusion, if the extra extinction of $E_{B-V}\sim(0.37-0.57)$ mag is due to the ISM of the host galaxy, the intrinsic location of the host galaxy would significantly derivate from the main distribution (green points in Figure \ref{host}) in the color--magnitude diagram of galaxies, suggesting that the extra $E_{B-V}\sim(0.37-0.57)$ mag is more likely due to the circumburst medium of GRB~240825A. This result is also consistent with the evolution of the color excess during the afterglow decay. The latter also seems to arise more likely from the circumburst medium.

We further perform the SED fitting for the host galaxy using the Python package $\texttt{piXedfit}$ \citep{2021ApJS..254...15A} using the data from the SDSS, PanSTARRS, and HSC-SSP surveys discussed in Section \ref{observation}. 
$\texttt{piXedfit}$ can be used to analyze spatially resolved properties of galaxies using multiband imaging data or in combination with integral field spectroscopic data. 
Since there is no built-in PanSTARRS filter information (e.g.,  central wavelengths, transmission efficiencies, etc.) in $\texttt{piXedfit}$, we first add its information to $\texttt{piXedfit}$, taking from the Filter Profile Service\footnote{\href{http://svo2.cab.inta-csic.es/theory/fps/}{http://svo2.cab.inta-csic.es/theory/fps/}} \citep{2020sea..confE.182R,2012ivoa.rept.1015R}. We generated 200,000 SED models at the rest-frame of the host galaxy, assuming the initial mass function of \cite{2003PASP..115..763C}, the double power-law form of the star formation history\footnote{We use the star formation history with a double power-law form of ${\rm SFR}\propto[(t/\tau)^{\alpha}+(t/\tau)^{-\beta}]^{-1}$, and the fitting results of $\alpha$, $\beta$ and $\tau$ are shown Figure \ref{cornerfigure}.} \citep{2018MNRAS.480.4379C,2013ApJ...770...57B,2017ApJ...839...26D,2021ApJS..254...15A}, and the dust attenuation law of \cite{2000ApJ...533..682C}. Furthermore, in the SED models, we include the nebular \citep{2017ApJ...840...44B,Ferland_1998,2013RMxAA..49..137F} and dust emissions \citep{2008MNRAS.388.1595D}. Applying the Bayesian inference with the MCMC method and setting the redshift as $z=0.695$, \citep{2024GCN.37293....1M}, we use the 200,000 SED models to fit the SED of the host galaxy, composed of the data from the SDSS, PanSTARRS, and HSC-SSP survey using the $\texttt{singleSEDfit}$ module. The IGM absorption model of \cite{2014MNRAS.442.1805I} is considered in this fitting process.

\begin{figure}
\centering
\includegraphics[width = 1\linewidth, trim = 0 0 0 0, clip]{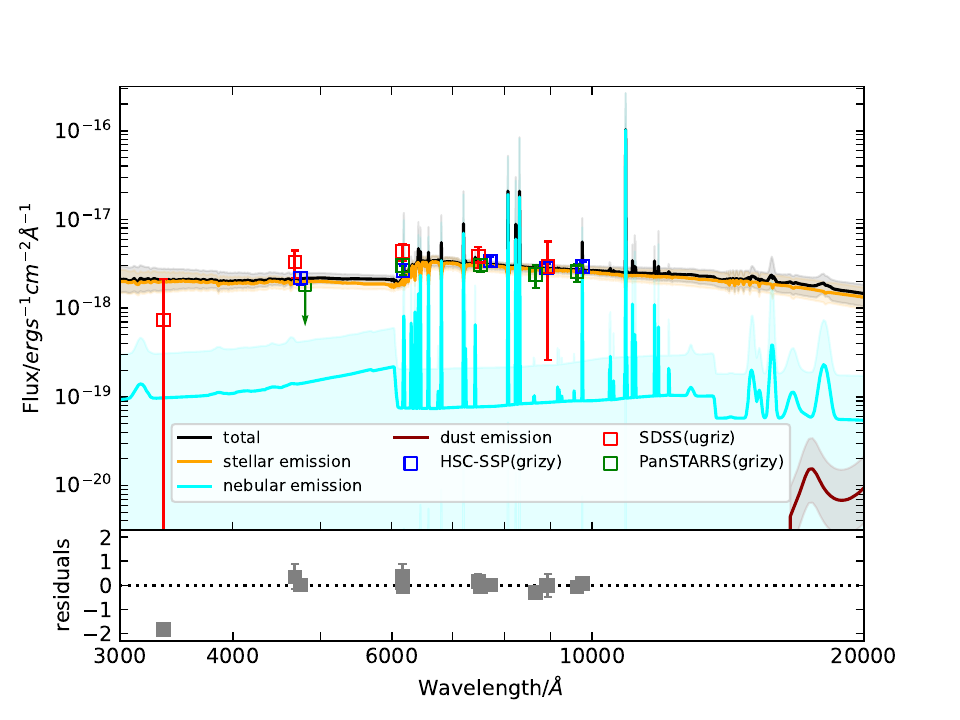}
\caption{The SED fitting of the GRB host galaxy using the Python package $\texttt{piXedfit}$ \citep{2021ApJS..254...15A}. The solid lines and their corresponding shaded area represent median posteriors and 1$\sigma$ uncertainties, respectively. The total emission is decomposed into stellar (orange), nebular (cyan), and dust (dark red) emissions. The blue, red, and green squares indicate the observed data from the HSC-SSP, SDSS, and PanSTARRS surveys, respectively. 
}\label{corner_sdss_panstarrs}
\end{figure}

\begin{figure*}
\centering
\includegraphics[width = 1.0\linewidth, trim = 0 0 0 0, clip]{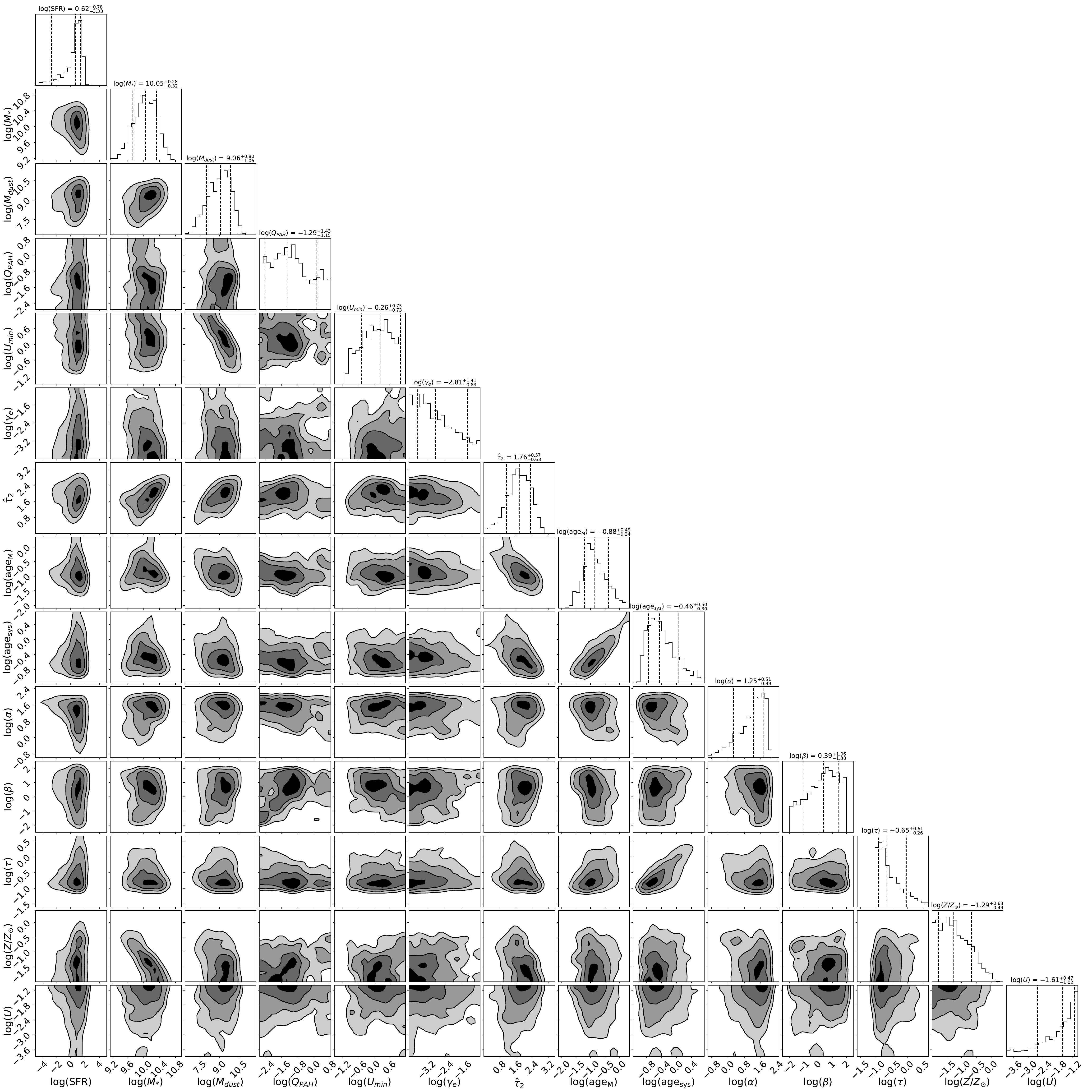}
\caption{Two-dimensional projections of the posterior probability distributions of the best-fit parameters of the host galaxy using the Python package $\texttt{piXedfit}$ \citep{2021ApJS..254...15A}. There are 14 parameters involved in our SED fitting as described in Table 1 of \cite{2021ApJS..254...15A}.}\label{cornerfigure}
\end{figure*}

The result and the two-dimensional projections of the posterior probability distributions of the best-fitting parameters are shown in Figure \ref{corner_sdss_panstarrs} and Figure \ref{cornerfigure}.
Contributions from the stellar, nebular, and dust emissions are all included in the galaxy SED model. We find that the $u$-band measurement from the SDSS has a quite large uncertainty and is likely to be a faint limit. The derived parameters of the host galaxy are: star formation rate $\log({\rm SFR}/M_{\odot}{\rm yr}^{-1})=0.6^{+0.8}_{-3.3}$, stellar mass $\log(M_*/M_{\odot})=10.0^{+0.3}_{-0.3}$, stellar metallicity $\log(Z/Z_{\odot})=-1.3^{+0.6}_{-0.5}$, evolving age of the stellar population\footnote{Notice that the evolving age of the stellar population refers to the age of a galaxy system, which represents the cosmic time since the onset of star formation and corresponds to the maximum age of stars within the system. It is different from the mass-weighted age of stars in the system, which serves as a representative age for the entire stellar population.} $\log({\rm age_{sys}}/{\rm Gyr})=-0.5^{+0.5}_{-0.3}$. We find that the results of the SED fitting are roughly consistent with the star formation rate and the stellar mass derived from the color-magnitude diagram discussed above.
However, due to the limited photometric data of the host galaxy and the lack of spectroscopic information, the results have relativity large uncertainties. 
Since the emissions of the massive stars in the star formation region are mainly in the blue band, the measurement of the SFR has a large uncertainty due to the large uncertainty of SDSS in its $u$ band.

\section{Discussions and Conclusions}\label{conclusion}

GRB~240825A was recently triggered by \texttt{Swift}--BAT at 15:52:59 UT on Aug. 25, 2024. In this work, we present the simultaneous multiband photometry of its early optical afterglow obtained with the Mephisto telescope which is capable of simultaneously imaging the same patch of sky in three bands. 
Mephisto observations began 128 seconds after the \texttt{Swift}--BAT trigger, and the light curves in the $uvgriz$ bands follow a single power-law decay. This early set of observations provides us with a strong constraint on the onset of the afterglow at the blastwave deceleration radius as well as on the initial Lorentz factor $\Gamma_0$ of the GRB ejecta. For two scenarios of the circumburst medium being composed of the general ISM and of the stellar winds, we find $\Gamma_0\gtrsim 139~{\rm and}~52$, respectively.

Based on the measured multiband light curves in $uvgriz$ bands, we find that the decay of the flux density of the early optical afterglow satisfies $F_{\nu,{\rm obs}}\propto t^{-\alpha_{\rm obs}}\nu^{-\beta_{\rm obs}}$ with a temporal decay index of $\alpha_{\rm obs}=1.340\pm0.002$ and a spectral index of $\beta_{\rm obs}=2.477\pm0.006$. It is noteworthy that the observed SED of GRB 240825A is much softer than those of most other GRBs \citep[e.g.,][]{Kann10,Kann11}. 
In particular, in the golden sample comprising 48 GRBs of \citet{Kann10}, there are only two sources (i.e., GRB 070802 and GRB 080310) with spectral indexes larger than $\beta_{\rm obs}\sim2$ and most GRBs have $\beta_{\rm obs}\sim1$ with uncertainties of $0.01-0.1$. 
This interesting discovery of the very soft SED of GRB 240825A may be attributed to the simultaneous photometry in the wide $uvgriz$ bands of Mephisto, especially for the contribution from the blue bands. And a relatively low uncertainty of the spectral index is obtained due to the same reason.

Assuming that the measured temporal decay index is intrinsic, $\alpha=\alpha_{\rm obs}=1.340$, there are only three main possible scenarios for the standard afterglow model to explain the feature of the optical afterglow:

1) Case I: ISM slow cooling, where $\nu_a<\min(\nu_m,\nu_c)$ with $\nu_m<\nu<\nu_c$, or $\nu_m<\nu_a<\nu_c$ with $\nu_a<\nu<\nu_c$. In this case, an electron spectral index of $p=2.79$ is required and the intrinsic spectral index directly inferred from $\alpha_{\rm obs}$ is $\beta(\alpha_{\rm obs})=0.89$.

2) Case II: wind slow cooling, where $\nu_a<\min(\nu_m,\nu_c)$ with $\nu_m<\nu<\nu_c$, or $\nu_m<\nu_a<\nu_c$ with $\nu_a<\nu<\nu_c$. In this case one has $p=2.12$ and $\beta(\alpha_{\rm obs})=0.56$. 

3) Case III: ISM/wind slow/fast cooling, where $\nu>\max(\nu_c,\nu_m)$. In this case one has $p=2.45$ and $\beta(\alpha_{\rm obs})=1.23$. 

The above results show that our measured spectral index of $\beta_{\rm obs}=2.477$ is much larger than those directly inferred from $\alpha_{\rm obs}$ in the standard afterglow model.
The result implies that the blue parts of the optical afterglow of GRB~240825A might be partially absorbed by the circumburst medium, leading to an extra extinction.
We then further check whether the extinction evolves with the afterglow decay. Using the afterglow temporal power-law decay to fit independently the light curves in different bands, we infer the magnitudes at different epochs via interpolation. 
Benefiting from the advantage of the simultaneous imaging in three channels of Mephisto, we find that the SED of the optical afterglow hardens as the afterglow decays, with the inferred color excess $E_{B-V}$ decreasing by $\sim0.26$ mag from 100 seconds 3000 seconds. This result suggests that the dust in the cirbumburst medium of GRB 240825A evolves as the afterglow decays. 

Another possibility of the SED hardening shown in the top panel of Figure \ref{multised} is due to the passage of a synchrotron characteristic frequency (e.g., $\nu_m$ or $\nu_c$) within the optical wavelength range. However, such a possibility is disfavored due to the following reasons. The observed very soft SED implies that it should be near the last segment of the spectrum of synchrotron radiation \citep[e.g.,][]{Sari98}. According to the standard afterglow model, the difference between the spectral indexes of the last two segments has two cases: $\Delta\beta=1/2$ or $\Delta\beta=(p-1)/2$ \citep[e.g.,][]{Sari98,Gao13,Zhang18}. However, the maximum difference of the observed spectral indexes in the first two days reaches $\Delta\beta_{\rm obs}\gtrsim1$, which is significantly larger than the predicted $\Delta\beta$ without the extinction. Therefore, the observed SED hardening seems not likely attributed to the passage of a synchrotron characteristic frequency. On the other hand, since the very soft SED directly reveals the existence of the extinction, it is more natural to attribute the SED hardening to the extinction evolution based on the principle of Occam's razor.

Since long GRBs have been proposed to occur in star-forming dusty regions of a galaxy \citep{Paczynski98}, the effects of dust extinction are important for understanding the environments and progenitors of long GRBs. In fact, only about 60\% of \texttt{Swift} GRBs are detected in optical wavelengths, and the remaining 40\% (so-called ``dark GRBs'') is undetectable in the optical bands, as a result of the attenuation of optical photons by the intervening dust \citep[e.g.,][]{Draine02}. Some evidence shows that the extinction curve toward the individual GRBs could be described by an SMC-like curve with a steep far-UV rise, implying the predominance of small grains in the local environments near the GRB sources \citep[e.g.,][]{Heintz17,Zafar18}.
Besides, some early optical afterglows of GRBs, e.g, GRB 111209A \citep{Stratta13} and GRB 120119A \citep{Morgan14}, show significant red-to-blue color change within a few hundred seconds after the prompt emission phase. Such a red-to-blue change also appears in our observations of the early afterglow of GRB~240825A, see the discussion in Section \ref{oa} shown. 
Thus, GRBs are good probes of their immediate environment, and the blue optical bands carry information about extinction, probing the existence of dust in the GRB emission region.
In physics, \citet{Hoang20} studied rotational disruption and alignment of dust grains by radiative torques induced by afterglows and found that optical-near-infrared extinction decreases and ultraviolet extinction increases due to the enhancement of small grains. This mechanism might explain the feature of the afterglow of GRB 240825A, see Section 3.3 of \citet{Hoang20} for a detailed discussion.

Finally, we also analyze the properties of the host galaxy of GRB~240825A, based on the data from the SDSS, PanSTARRS and HSC-SSP surveys. Using the SED models of galaxies \citep{2021ApJS..254...15A}, the star formation rate, the stellar mass, the stellar metallicity and the stellar evolving age of the stellar population of the host GRB galaxy are estimated to be $\log({\rm SFR}/M_{\odot}{\rm yr}^{-1})=0.6^{+0.8}_{-3.3}$, $\log(M_*/M_{\odot})=10.0^{+0.3}_{-0.3}$,  $\log(Z/Z_{\odot})=-1.3^{+0.6}_{-0.5}$, $\log({\rm age_{sys}}/{\rm Gyr})=-0.5^{+0.5}_{-0.3}$, respectively, pointing to a gas-rich, star-forming, medium-size galaxy. 
\citet{Savaglio09} presented the properties of the host galaxies of the sample including 46 GRBs. The average stellar mass is $10^{9.3}M_\odot$ with 1$\sigma$ dispersion of 0.8 dex. The average metallicity for a subsample is about $1/6$ of the solar's. The star formation rate is $(0.01-36)M_\odot{\rm yr^{-1}}$.
It is worth noting that GRB hosts are smaller in size and they have higher stellar mass and star formation rate
surface densities than field galaxies \citep{Schneider22}, and almost $\sim20\%$ of GRBs are heavily dust-obscured \citep{Perley16a,Perley16b}.
For the host galaxy of GRB 240825A, its properties are relatively moderate in the GRB sample but with relatively low metallicity. Although its optical afterglow shows a significant extinction, both the evolving properties of the afterglow and the host location in the color-magnitude diagram suggest that its extinction is more likely from the circumburst medium.


\section*{Acknowledgments}

Mephisto is developed at and operated by the South-Western Institute for Astronomy Research of Yunnan University (SWIFAR-YNU), funded by the ``Yunnan University Development Plan for World-Class University'' and ``Yunnan University Development Plan for World-Class Astronomy Discipline''. The authors acknowledge supports from the ``Science \& Technology Champion Project'' (202005AB160002) and from two ``Team Projects'' -- the ``Innovation Team'' (202105AE160021) and the ``Top Team'' (202305AT350002), all funded by the ``Yunnan Revitalization Talent Support Program''.
Y.P.Y. is supported by the National Key Research and Development Program of China (2024YFA1611603), the National Natural Science Foundation of China grant No.12473047 and the National SKA Program of China (2022SKA0130100). J.Z. acknowledges financial support from NSFC grant No. 12103063.
X. H. and P. Z. are supported by the Strategic Priority Research Program of the Chinese Academy of Sciences, Grant No. XDB0550101.
L. X. is supported by the Strategic Priority Research Program of the Chinese Academy of Sciences, Grant No. XDB0550401.
C. W. is supported by the SVOM project, a mission in the Strategic Priority Program on Space Science of CAS.
We thank the anonymous referee for providing helpful comments and suggestions.
We also acknowledge the helpful discussions with Bing Zhang, Yi-Zhong Fan, Ye Li, and Yun Wang.

The Pan-STARRS1 Surveys (PS1) and the PS1 public science archive have been made possible through contributions by the Institute for Astronomy, the University of Hawaii, the Pan-STARRS Project Office, the Max-Planck Society and its participating institutes, the Max Planck Institute for Astronomy, Heidelberg and the Max Planck Institute for Extraterrestrial Physics, Garching, The Johns Hopkins University, Durham University, the University of Edinburgh, the Queen's University Belfast, the Harvard-Smithsonian Center for Astrophysics, the Las Cumbres Observatory Global Telescope Network Incorporated, the National Central University of Taiwan, the Space Telescope Science Institute, the National Aeronautics and Space Administration under Grant No. NNX08AR22G issued through the Planetary Science Division of the NASA Science Mission Directorate, the National Science Foundation Grant No. AST-1238877, the University of Maryland, Eotvos Lorand University (ELTE), the Los Alamos National Laboratory, and the Gordon and Betty Moore Foundation.

The Hyper Suprime-Cam (HSC) collaboration includes the astronomical communities of Japan and Taiwan, and Princeton University. The HSC instrumentation and software were developed by the National Astronomical Observatory of Japan (NAOJ), the Kavli Institute for the Physics and Mathematics of the Universe (Kavli IPMU), the University of Tokyo, the High Energy Accelerator Research Organization (KEK), the Academia Sinica Institute for Astronomy and Astrophysics in Taiwan (ASIAA), and Princeton University. Funding was contributed by the FIRST program from the Japanese Cabinet Office, the Ministry of Education, Culture, Sports, Science and Technology (MEXT), the Japan Society for the Promotion of Science (JSPS), Japan Science and Technology Agency (JST), the Toray Science Foundation, NAOJ, Kavli IPMU, KEK, ASIAA, and Princeton University. 

This paper makes use of software developed for Vera C. Rubin Observatory. We thank the Rubin Observatory for making their code available as free software at \href{http://pipelines.lsst.io/}{http://pipelines.lsst.io/}.

This paper is based on data collected at the Subaru Telescope and retrieved from the HSC data archive system, which is operated by the Subaru Telescope and Astronomy Data Center (ADC) at NAOJ. Data analysis was in part carried out with the cooperation of Center for Computational Astrophysics (CfCA), NAOJ. We are honored and grateful for the opportunity of observing the Universe from Maunakea, which has the cultural, historical and natural significance in Hawaii.

Funding for the Sloan Digital Sky Survey IV has been provided by the Alfred P. Sloan Foundation, the U.S. Department of Energy Office of Science, and the Participating Institutions. SDSS-IV acknowledges support and resources from the Center for High Performance Computing at the University of Utah. The SDSS website is \href{www.sdss.org}{www.sdss.org}.

SDSS-IV is managed by the Astrophysical Research Consortium for the Participating Institutions of the SDSS Collaboration including the Brazilian Participation Group, the Carnegie Institution for Science, Carnegie Mellon University, Center for Astrophysics | Harvard \& Smithsonian, the Chilean Participation Group, the French Participation Group, Instituto de Astrof\'isica de Canarias, The Johns Hopkins University, Kavli Institute for the Physics and Mathematics of the Universe (IPMU) / University of Tokyo, the Korean Participation Group, Lawrence Berkeley National Laboratory, Leibniz Institut f\"ur Astrophysik Potsdam (AIP),  Max-Planck-Institut f\"ur Astronomie (MPIA Heidelberg), Max-Planck-Institut f\"ur Astrophysik (MPA Garching), Max-Planck-Institut f\"ur Extraterrestrische Physik (MPE), National Astronomical Observatories of China, New Mexico State University, New York University, University of Notre Dame, Observat\'ario Nacional / MCTI, The Ohio State University, Pennsylvania State University, Shanghai Astronomical Observatory, United Kingdom Participation Group, Universidad Nacional Aut\'onoma de M\'exico, University of Arizona, University of Colorado Boulder, University of Oxford, University of Portsmouth, University of Utah, University of Virginia, University of Washington, University of Wisconsin, Vanderbilt University, and Yale University.

\appendix

\begin{longtable}{ccccccc}
  \caption{Photometric data of GRB~240825A obtained with Mephisto in $ugi$ and $vrz$ bands. All measurements are in AB magnitudes and have been corrected for the Milky Way extinction. The times correspond to the middle-point of each exposure.
  Note that the uncertainties in the table mainly include the measurement uncertainties. The uncertainties of the photometric calibration are estimated to be better than 0.03, 0.01 and 0.005 mag for $u,v$, and $griz$, respectively.
} \label{tab}\\
  \toprule
  \textbf{MJD/day}  & \textbf{u/mag} & \textbf{v/mag} & \textbf{g/mag} & \textbf{r/mag} & \textbf{i/mag} & \textbf{z/mag} \\
  \midrule
  \endfirsthead
  \multicolumn{5}{c}{{\tablename\ \thetable{} -- continued}} \\
  \toprule
 \textbf{MJD/day}  & \textbf{u/mag} & \textbf{v/mag} & \textbf{g/mag}& \textbf{r/mag} & \textbf{i/mag} & \textbf{z/mag} \\
  \midrule
  \endhead
  \bottomrule
  \endfoot
  \bottomrule
  \endlastfoot
  
60547.66360&--&--&--&14.322${\pm}$0.002&--&--\\
60547.66395&--&--&--&--&--&13.539${\pm}$0.003\\
60547.66429&--&16.49${\pm}$0.01&--&--&--&--\\
60547.66446&--&--&--&15.036${\pm}$0.004&--&--\\
60547.66531&--&--&--&15.487${\pm}$0.005&--&--\\
60547.66559&--&--&--&--&--&14.619${\pm}$0.006\\
60547.66617&--&--&--&15.810${\pm}$0.007&--&--\\
60547.66654&--&17.45${\pm}$0.03&--&--&--&--\\
60547.66703&--&--&--&16.057${\pm}$0.008&--&--\\
60547.66721&--&--&--&--&--&15.143${\pm}$0.009\\
60547.66788&--&--&--&16.23${\pm}$0.01&--&--\\
60547.66874&--&--&--&16.44${\pm}$0.01&--&--\\
60547.66878&--&18.05${\pm}$0.04&--&--&--&--\\
60547.66884&--&--&--&--&--&15.52${\pm}$0.01\\
60547.66958&--&--&--&16.60${\pm}$0.01&--&--\\
60547.67044&--&--&--&16.67${\pm}$0.01&--&--\\
60547.67046&--&--&--&--&--&15.78${\pm}$0.01\\
60547.67130&--&--&--&16.80${\pm}$0.01&--&--\\
60547.67266&--&--&17.31${\pm}$0.03&--&--&--\\
60547.67301&--&--&--&--&16.33${\pm}$0.01&--\\
60547.67336&18.89${\pm}$0.09&--&--&--&--&--\\
60547.67352&--&--&17.34${\pm}$0.03&--&--&--\\
60547.67438&--&--&17.41${\pm}$0.03&--&--&--\\
60547.67464&--&--&--&--&16.48${\pm}$0.01&--\\
60547.67523&--&--&17.44${\pm}$0.04&--&--&--\\
60547.67560&18.6${\pm}$0.1&--&--&--&--&--\\
60547.67608&--&--&17.57${\pm}$0.04&--&--&--\\
60547.67627&--&--&--&--&16.63${\pm}$0.01&--\\
60547.67693&--&--&17.68${\pm}$0.04&--&--&--\\
60547.67779&--&--&17.64${\pm}$0.04&--&--&--\\
60547.67785&19.2${\pm}$0.2&--&--&--&--&--\\
60547.67789&--&--&--&--&16.74${\pm}$0.02&--\\
60547.67865&--&--&17.77${\pm}$0.04&--&--&--\\
60547.67949&--&--&17.66${\pm}$0.06&--&--&--\\
60547.67951&--&--&--&--&16.81${\pm}$0.02&--\\
60547.68035&--&--&18.0${\pm}$0.1&--&--&--\\
60547.68502&--&--&18.01${\pm}$0.06&--&--&--\\
60547.68537&--&--&--&--&17.13${\pm}$0.02&--\\
60547.68572&19.7${\pm}$0.2&--&--&--&--&--\\
60547.68588&--&--&18.28${\pm}$0.06&--&--&--\\
60547.68674&--&--&18.15${\pm}$0.06&--&--&--\\
60547.68731&--&--&--&--&17.18${\pm}$0.02&--\\
60547.68759&--&--&18.24${\pm}$0.06&--&--&--\\
60547.68796&19.1${\pm}$0.2&--&--&--&--&--\\
60547.68844&--&--&18.22${\pm}$0.06&--&--&--\\
60547.68892&--&--&--&--&17.25${\pm}$0.02&--\\
60547.68929&--&--&18.22${\pm}$0.06&--&--&--\\
60547.69015&--&--&18.42${\pm}$0.07&--&--&--\\
60547.69021&20.0${\pm}$0.2&--&--&--&--&--\\
60547.69054&--&--&--&--&17.27${\pm}$0.02&--\\
60547.69101&--&--&18.43${\pm}$0.07&--&--&--\\
60547.69186&--&--&18.35${\pm}$0.07&--&--&--\\
60547.69216&--&--&--&--&17.35${\pm}$0.02&--\\
60547.69271&--&--&18.45${\pm}$0.07&--&--&--\\
60547.69686&--&--&--&18.24${\pm}$0.07&--&--\\
60547.69721&--&--&--&--&--&17.41${\pm}$0.07\\
60547.69756&--&19.6${\pm}$0.2&--&--&--&--\\
60547.69772&--&--&--&18.3${\pm}$0.1&--&--\\
60547.69858&--&--&--&18.3${\pm}$0.2&--&--\\
60547.69889&--&--&--&--&--&17.5${\pm}$0.1\\
60547.69942&--&--&--&18.33${\pm}$0.09&--&--\\
60547.69980&--&20.7${\pm}$0.4&--&--&--&--\\
60547.70028&--&--&--&18.35${\pm}$0.06&--&--\\
60547.70051&--&--&--&--&--&17.48${\pm}$0.05\\
60547.70113&--&--&--&18.46${\pm}$0.06&--&--\\
60547.70199&--&--&--&18.32${\pm}$0.05&--&--\\
60547.70205&--&19.9${\pm}$0.2&--&--&--&--\\
60547.70212&--&--&--&--&--&17.45${\pm}$0.05\\
60547.70285&--&--&--&18.36${\pm}$0.05&--&--\\
60547.70369&--&--&--&18.35${\pm}$0.05&--&--\\
60547.70374&--&--&--&--&--&17.50${\pm}$0.05\\
60547.70455&--&--&--&18.39${\pm}$0.05&--&--\\
60548.63584&--&--&--&21.7${\pm}$0.1&--&--\\
60548.63920&--&--&22.1${\pm}$0.1&--&--&--\\
60548.66045&--&--&--&--&20.9${\pm}$0.1&--\\
60549.65264&--&--&22.4${\pm}$0.1&--&--&--\\
60549.65272&--&--&--&--&21.3${\pm}$0.5&--\\
60549.66503&--&--&--&21.9${\pm}$0.1&--&--\\
60549.66509&--&--&--&--&--&21.2${\pm}$0.5\\
60550.67669&--&--&--&--&21.2${\pm}$0.1&--\\
60550.68991&--&--&--&22.6${\pm}$0.2&--&--\\
60550.68998&--&--&--&--&--&21.5${\pm}$0.3\\

\hline
\end{longtable}


\end{document}